\documentclass[runningheads]{llncs}
\usepackage[T1]{fontenc}
\usepackage{cite}
\usepackage{amsmath,amssymb,amsfonts,bm}
\usepackage{footmisc}
\usepackage{algorithmic}
\usepackage[colorlinks=true,linkcolor=blue,citecolor=blue,urlcolor=blue]{hyperref}
\usepackage{graphicx}
\usepackage{caption}
\usepackage{subcaption}
\usepackage{textcomp}
\usepackage{wrapfig}
\usepackage{tablefootnote}
\usepackage{comment}
\usepackage{xcolor}
\usepackage{url}
\usepackage{siunitx}
\usepackage{booktabs}
\usepackage{multirow}
\usepackage{lmodern}
\usepackage{array}\usepackage{makecell}

\ifcsundef{fact}{%
}

\begin{document}

\title{MP-RW-LSH: An Efficient Multi-Probe LSH Solution to ANNS in $L_1$ Distance}
\titlerunning{An Efficient Multi-Probe LSH Solution to ANNS in $L_1$ Distance}
%
\author{Huayi Wang\inst{1} \and Jingfan Meng\inst{1} \and 
Long Gong\inst{2} \and Jun Xu\inst{1} \and Mitsunori Ogihara\inst{3}}
\authorrunning{H. Wang et al.}
%

\institute{Georgia Institute of Technology, USA \\
\email{\{huayiwang,jmeng40\}@gatech.edu jx@cc.gatech.edu} \and
Facebook, USA \\
\email{lgong30@fb.com}\\ \and
University of Miami, USA\\
\email{m.ogihara@miami.edu}}


\maketitle

\begin{abstract}
Approximate Nearest Neighbor Search (ANNS) is a fundamental algorithmic problem, with numerous applications in
many areas of computer science. Locality-sensitive hashing (LSH) is one of the most popular solution approaches for ANNS. 
A common shortcoming of many LSH schemes is that since they probe only a single bucket in a hash table, they 
need to use a large number of hash tables to achieve a high 
query accuracy.  For ANNS-$L_2$, a multi-probe scheme was proposed to overcome this drawback by strategically probing multiple buckets in a hash table.
In this work, we propose MP-RW-LSH, the first and so far only multi-probe LSH solution to ANNS in $L_1$ distance.  
Another contribution of this work is to explain why a state-of-the-art ANNS-$L_1$ solution called Cauchy projection LSH (CP-LSH) is 
fundamentally not suitable for 
multi-probe extension. 
We show that MP-RW-LSH
uses 15 to 53 times fewer hash tables than CP-LSH for achieving similar query accuracies. 
\end{abstract}

\section{Introduction}

Approximate Nearest Neighbor Search (ANNS) is a fundamental algorithmic problem, with numerous applications
in many areas of computer science, including
informational retrieval~\cite{Lin_retrieval_2015},
recommendations~\cite{Qi_Recommendation_2017},
near-duplication detections~\cite{Sood_duplicateDetection_2011}, etc.  
 In ANNS, given a query point $\mathbf{q}$, 
we search in a massive dataset $\mathcal{D}$, that lies in a high-dimensional space,  
for one or more points in $\mathcal{D}$ that are \emph{among the closest} to $\mathbf{q}$ according to some
distance metric.  Throughout this paper, like in the case of $\mathbf{q}$, all letters in bold represent vectors.   

The ANNS literature is mostly focused on ANNS in the Euclidean ($L_2$) distance, or ANNS-$L_2$ in short.  
In this work, we focus instead on ANNS-$L_1$, ANNS in the Manhattan ($L_1$) distance, which is much less thoroughly studied in comparison.
ANNS-$L_1$ is nonetheless an extremely important problem for two reasons.  First, it arises in almost all application domains of ANNS-$L_2$.
Second, set and multiset 
similarity search and join~\cite{satuluri2012bayesian,Zhang_EmbedJoinEfficientEdit_2017}, 
an increasingly important family of ANNS problems that are widely used in database application domains such as data cleaning\cite{chaudhuri2006primitive}, social network mining\cite{spertus2005evaluating} and information retrieval~\cite{wang2007topical}, can be reduced to ANNS-$L_1$~\cite{gong2020idec}; hence any breakthrough on the latter is automatically one on the former. 

One of the most popular ANNS solution approaches is Locality-Sensitive Hashing (LSH)~\cite{Indyk_ANN_1998}. 
The key intellectual component of an LSH scheme is its hash function family 
$\mathcal{H}$.
Any function $h$ sampled uniformly at random from $\mathcal{H}$ has the following 
nice collision property:  It maps two distinct points in $\mathcal{D}$ to the same hash value with probability $p_1$
if they are close to each other (say no more than distance $r_1$ apart) and with probability $p_2 < p_1$ if they are far apart (say more than $r_2 > r_1$ apart), respectively.
Such an LSH scheme can achieve a query time complexity of roughly $O(n^\rho)$,
where $\rho \triangleq \frac{\log(1/p_1)}{\log(1/p_2)}$ is called the \emph{quality} of the LSH family, and $n$ is the number of points in $\mathcal{D}$.
However, an LSH scheme requires the maintenance and search of
a large number ($O(n^\rho)$ in theory~\cite{Indyk_ANN_1998} and tens to hundreds in practice~\cite{buhler2001efficient}) of hash tables, for the
reason explained next.

Whenever possible, in the rest of this paper, we focus on only one of these hash tables and explain how it is probed for the 
nearest neighbors of $\mathbf{q}$.  
In this hash table, an LSH scheme 
probes only a single bucket that has the highest success probability (of containing a nearest neighbor of $\mathbf{q}$):  $\mathbf{h}(\mathbf{q})$, the bucket
that $\mathbf{q}$ is hashed to by an LSH function vector $\mathbf{h}(\cdot) \triangleq \langle h_1(\cdot), \cdots, h_M(\cdot)\rangle$.
We refer to $\mathbf{h}(\mathbf{q})$ as the {\it epicenter bucket} in the sequel.  
However, the success probability of the epicenter bucket is still quite low, since this probability decays exponentially with $M$, 
and M can be as large as 20 in some LSH schemes. 
Hence a large number of hash tables have to be used to 
boost the overall success probability of finding at least one nearest neighbor.  



Multi-probe~\cite{lv2007multi} was proposed  
for boosting this success probability when the 
Gaussian-projection LSH scheme (GP-LSH)~\cite{Datar_LocalitysensitiveHashing_2004} for  
ANNS-$L_2$ is used as the baseline LSH.
The idea of multi-probe is that, the algorithm probes not only the epicenter bucket, but also 
$T > 0$ other nearby buckets whose success probabilities are among the 
$T+1$ highest.
This way, the total success probability
can be significantly
increased, and the number of hash tables used for reaching a target success probability can be significantly
reduced.  

With this multi-probe enhancement, the resulting LSH scheme, which we call MP-GP-LSH, can 
significantly reduce the number of hash tables needed 
while achieving 
a similar query accuracy and query time as GP-LSH, its baseline LSH.  
Due to its spectacular efficacy, MP-GP-LSH has since been deployed in 
various systems including smartphone applications~\cite{rublee2011orb}, audio content retrieval~\cite{yu2010combining}, automatic product suggestions~\cite{kalantidis2013getting}, etc.
We will explain in Sect.\,\ref{subsec:mplsh} that the efficacy of MP-GP-LSH
stems entirely from the following property of GP-LSH:
The success probability of a bucket 
decreases roughly at the ``Gaussian pdf rate'' $O(e^{-c{d_2}^2})$, where $d_2$ is the bucket's $L_2$ distance from the 
epicenter (to be defined in Sect.\,\ref{subsec:mplsh}), and $c > 0$ is a constant.
In comparison, it appears hard to apply the multi-probe approach to any other LSH scheme that does not have this property.
Currently multi-probe LSH solutions exist only for ANNS in the Chi-squared distance~\cite{gorisse2011locality}
and in the angular distance~\cite{andoni2015practical}.

In this work, we propose MP-RW-LSH, the first and so far only multi-probe LSH solution for ANNS-$L_1$.
Our solution significantly outperforms 
Cauchy projection LSH (CP-LSH)~\cite{Datar_LocalitysensitiveHashing_2004}, the  
state-of-the-art LSH scheme for ANNS-$L_1$.  Our solution however is not a multi-probe extension of CP-LSH.  
In fact, we discover that CP-LSH is fundamentally not suitable for multi-probe for the following reason:  The total success probability of the top-$(T+1)$ buckets
remains quite low even when $T$ is very large thanks to
the heavy-tail nature~\cite{resnick2007heavy} of its underlying Cauchy distribution.

We propose a new LSH scheme for ANNS-$L_1$ that is much better suited for multi-probe.
We call it random-walk LSH (RW-LSH), because any raw hash value function (defined later) $f$ in it has the following property:  
Given any two nonnegative integer data points $\mathbf{s}$ and $\mathbf{t}$, 
$f(\mathbf{s}) - f(\mathbf{t})$, the difference between their raw hash values, has the same probability
distribution as that of a $d_1$-step random walk, where $d_1 = \|\mathbf{s} - \mathbf{t}\|_1$ is their $L_1$ distance.
Hence, when $d_1$ is large, this difference converges to
a zero-mean Gaussian distribution with variance $d_1$.   
As a result, given a query point $\mathbf{q}$, the success probability of a bucket
decays in the same aforementioned Gaussian pdf manner as in GP-LSH.  
Hence RW-LSH can be extended to MP-RW-LSH in almost the same way as GP-LSH (to MP-GP-LSH). 
We will show that MP-RW-LSH strikes a much better tradeoff between scalability and query efficiency than 
CP-LSH~\cite{Datar_LocalitysensitiveHashing_2004} and SRS~\cite{Sun16_thesis}, which are the two state-of-the-art LSH-based ANNS-$L_1$ solutions in terms of 
query efficiency and of scalability respectively.

\section{Preliminaries} \label{sec:background}

\subsection{Locality-Sensitive Hashing} \label{subsec:lsh_background}
In an LSH scheme, typically a vector of $M > 1$ LSH functions $\mathbf{h} = \langle h_1$, $h_2$, $\cdots$, $h_M\rangle$ 
are used to map each point $\mathbf{s}$ in $\mathcal{D}$ to an $M$-dimensional vector of hash values 
$\mathbf{h}( \mathbf{s}) = \langle h_1(\mathbf{s}), h_2(\mathbf{s}), \cdots, h_M(\mathbf{s})\rangle$.  
This point $\mathbf{s}$ is to be stored in a hash bucket indexed 
by the vector $\mathbf{h}(\mathbf{s})$;
hence we identify this hash bucket as $\mathbf{h}(\mathbf{s})$. 
Then given a query point $\mathbf{q}$, the search procedure is to
probe all points in the hash bucket $\mathbf{h}(\mathbf{q})$
in the hope that 
some nearest neighbors of $\mathbf{q}$ are mapped to the same vector (bucket). 

Let $p_1$ and $p_2$ be as defined above.  
Given a query point $\mathbf{q}$, the number of \textit{spurious points} in $\mathcal{D}$ (say containing $n$ points), is 
equal to $p_2^M n$ in expectation, 
where a spurious point is one that is 
mapped by the $M$ LSH functions to the same vector as, but is not actually close to, 
the query point $\mathbf{q}$.   
Since this number, which corresponds to the time cost of probing each bucket, needs to be kept low at $O(1)$,
we need $M = \log_{1/p_2}n + O(1)$ LSH functions.  However, in this case
the probability with which any good point (one that is close to $\mathbf{q}$) is hashed to the epicenter bucket, is only 
$p_1^M = O(n^{-\rho})$, where 
$\rho =\frac{\log 1/p_1}{\log 1/ p_2}$ is the quality of the LSH family as defined above.  
Hence roughly 
$O(n^{\rho})$ hash tables have to be used to guarantee that any good point has a probability at least $1-e^{-1}$ to be found in at least one hash table. 
Therefore, the query time complexity of such an LSH scheme is also $O(n^\rho)$.


We now describe such an LSH function $h_i$ (in the vector $\mathbf{h}$ defined above) used in the three aforementioned LSH schemes: GP-LSH, CP-LSH, and RW-LSH.
In all these three LSH schemes, $h_i$ takes the same following form:  $h_i(\mathbf{s}) =  \left\lfloor \frac{f_i(\mathbf{s}) + b_i}{W} \right\rfloor$, 
where $W > 0$ is a constant and $b_i$ is a random variable (fixed after generation) uniformly distributed in $[0, W]$.
Here $f_i(\mathbf{s})$ is called the raw hash value of $\mathbf{s}$. Clearly, each bucket corresponds to an $M$-dimensional
cube with width $W$ in each dimension; any point $\mathbf{s}$ 
whose raw hash value vector $\mathbf{f}(\mathbf{s}) = \langle f_1(\mathbf{s}),f_2(\mathbf{s}),\cdots,f_M(\mathbf{s})\rangle$ falls into this cube belongs to the corresponding hash bucket $\mathbf{h}(\mathbf{s})$.
Given a query point $\mathbf{q}$, we 
refer to its shifted raw hash value vector $\mathbf{f}(\mathbf{q})+\mathbf{b}$ as the {\it epicenter} and its hash bucket $\mathbf{h}(\mathbf{q})$ as the 
{\it epicenter bucket} in the sequel.

For an $m$-dimensional point $\mathbf{s} = \langle s_1,s_2,\cdots,s_m \rangle$, the raw hash value function $f_i(\cdot)$ takes the same form in GP-LSH and RW-LSH:  $f_i(\mathbf{s}) = \mathbf{s}\cdot \bm{\eta}$, 
where ``$\cdot$'' is the inner product (which is mathematically a {\it projection}). 
GP-LSH and CP-LSH differ only in the choice of $\bm{\eta}$.  
In GP-LSH, $\bm{\eta}$ is an $m$-dimensional i.i.d. standard Gaussian random vector (fixed after generation), so its $f_i$ is called a Gaussian projection.
In CP-LSH, $\bm{\eta}$ is an $m$-dimensional i.i.d. standard Cauchy random vector, so its $f_i$ is called a Cauchy projection.
In RW-LSH, $f_i$ takes a slightly different form that will be described in Sect.\,\ref{subsec:lsh-scheme}.
\subsection{Multi-Probe LSH} \label{subsec:mplsh}


In this section, we describe MP-GP-LSH,
the original multi-probe LSH scheme for ANNS-$L_2$~\cite{lv2007multi} that uses the Gaussian projection LSH (GP-LSH) as the 
baseline.
Again, we fix a query point $\mathbf{q}$ and one hash table, and focus on probing for nearest neighbors of $\mathbf{q}$ in this hash table.
As explained earlier, the idea of multi-probe is to probe the top-(T+1) buckets including the 
epicenter bucket $\mathbf{h}(\mathbf{q}) = \langle h_1(\mathbf{q}), h_2(\mathbf{q}), \cdots, h_M(\mathbf{q})\rangle$.
It can be shown that these buckets have to be in the same ``neighborhood'' as the epicenter bucket in the following sense:
For any such bucket $\bm{\beta} = \langle\beta_1, \beta_2, \cdots, \beta_M\rangle$, the value of its $i^{th}$ coordinate $\beta_i$ differs from $h_i(\mathbf{q})$, the $i^{th}$ coordinate of the epicenter bucket, by at most $1$. 
In other words, 
each $\beta_i$ takes one of the following three possible values:  $h_i(\mathbf{q}) - 1$, $h_i(\mathbf{q})$, and $h_i(\mathbf{q}) + 1$.  
We can represent each such bucket $\bm{\beta}$ by $\bm{\beta} - \mathbf{h}(\mathbf{q})$, its offset from $\mathbf{h}(\mathbf{q})$.  
This offset, denoted as $\bm{\delta} = \langle\delta_1, \delta_2, \cdots, \delta_M\rangle$ where $\delta_i \in \{-1, 0, 1\}$, is called the hash perturbation vector~\cite{lv2007multi}.

In a multi-probe LSH scheme, the top-(T+1) buckets need to be first identified and then probed in the decreasing order of their success probabilities;
we call this ordered list the \textit{optimal probing sequence}. 
However, computing the optimal probing sequence is not an easy undertaking in general.  A naive solution is to enumerate
every bucket in the ``neighborhood'' of $\mathbf{h}(\mathbf{q})$ and calculate its success probability.  
This algorithm is however very expensive computationally 
since there are $3^M-1$ buckets in the ``neighborhood''.
Hence, three refinements were proposed in~\cite{lv2007multi} for more efficiently computing the optimal or a near-optimal probing sequence. 

\noindent
\textbf{The First Refinement.} The first refinement is the following algorithm of ``winding down equi-height lines from the peak''.   
We can model the ``neighborhood'' of $\mathbf{h}(\mathbf{q})$ as an $M$-dimensional equi-height map in which the height of a point (bucket) is its success probability.  
Radiating from the peak is a series of equi-height ``lines'' ($(M-1)$-dimensional ``manifolds''); 
the exact shapes of these ``lines''
depend on the probability distributions induced by the LSH family. 
Under this modeling, intuitively an efficient algorithm for computing the optimal
probing sequence is to start at the peak (epicenter bucket) and then ``wind down'' one equi-height (equal success probability) ``line'' after next until the top-(T+1) 
are identified.  
It was shown in~\cite{lv2007multi} that this ``winding down'' algorithm can be implemented straightforwardly using a heap data structure with success probabilities as keys.  
This algorithm has a salient property:  It traverses, and computes the success probabilities of, at most $O(T)$ buckets, for identifying the top-(T+1) 
buckets.  This is a huge improvement over the naive algorithm that needs to traverse $3^M-1$ buckets.  
We refer to this refinement as the heap algorithm in the sequel.

We discover that the heap algorithm can find the optimal probing sequence for all baseline LSH schemes (when they are extended for 
multi-probe)
that satisfy the following property:  The equi-height map of the LSH scheme has no other peak (local maximum).  
We also find that the other two baseline LSH schemes used in this paper, namely CP-LSH and RW-LSH, both satisfy this property and can
use the heap algorithm for their respective multi-probe extensions.
However, the heap algorithm is still too computationally expensive for the following reason.  
For each bucket in the ``neighborhood'', its success probability is the product of $M$ different probability values (one corresponding to each hash function $h_i$
that is independent of others) 
each of which usually
takes a nontrivial amount of time to compute.
Hence, even to compute $O(T)$ such success probabilities
can be quite time-consuming.

\begin{wrapfigure}{R}{0.5\textwidth}
	\includegraphics[width=0.54\textwidth]{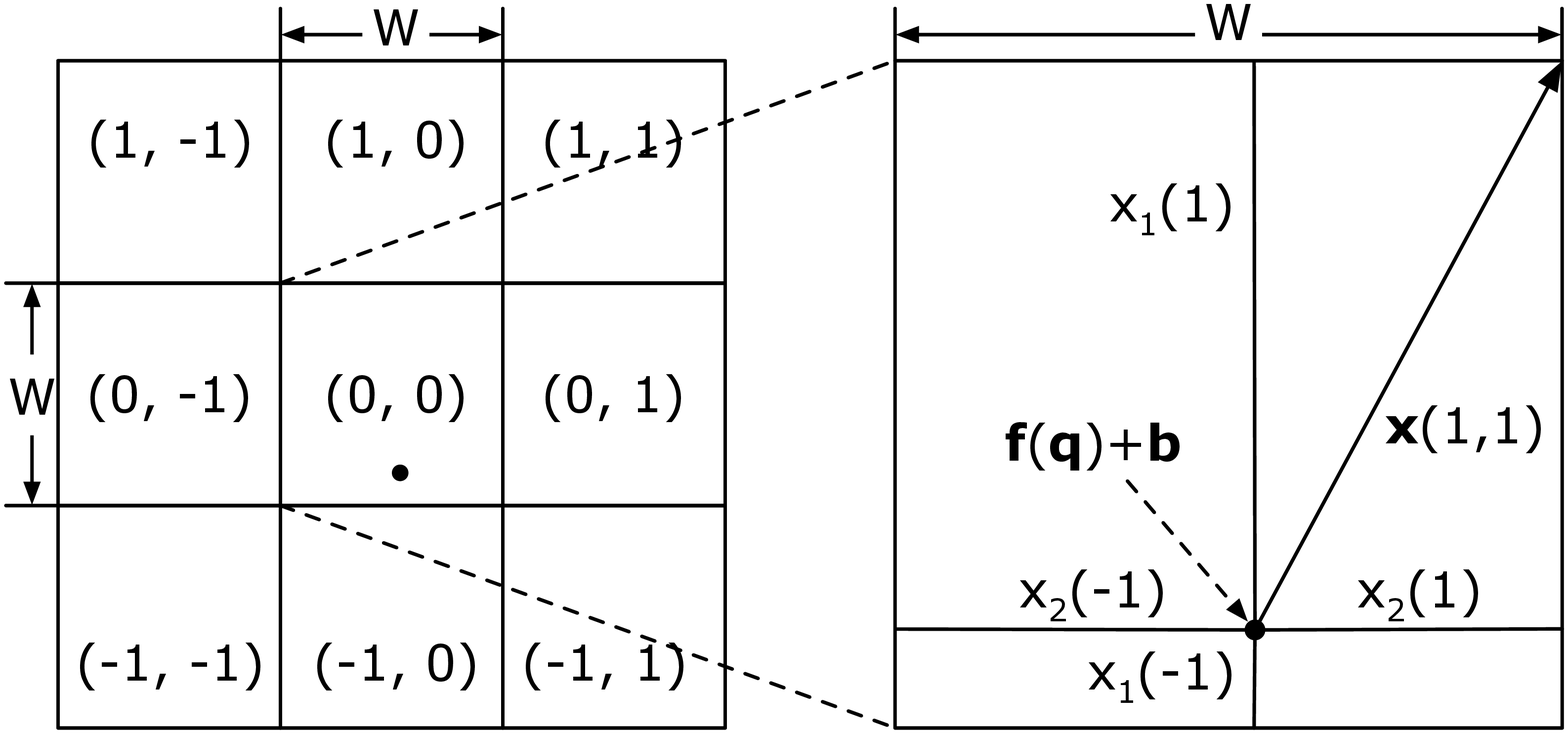}
	\caption{A toy example on multi-probe}\label{fig:multi_probe}
\end{wrapfigure}
\noindent
\textbf{The Second Refinement.} The objective of the second refinement is to significantly reduce the constant factor in this $O(T)$.  So far it works only for GP-LSH,
and it is not known whether it can be made to work for any other baseline LSH scheme.  
The second refinement is based on a critical insight:  
It is not the exact success probability values of these $O(T)$
buckets, but the relative order among these values that need to be determined.  
In the case of the GP-LSH, the relative order is much easier to compute than the probability values.

For ease of presentation, we simplify and introduce some notations.  
Recall that each $h_i(\mathbf{q}) = \lfloor \frac{f_i(\mathbf{q}) + b_i}{W}\rfloor$, where 
$f_i$ is a Gaussian projection and $b_i$ is uniformly distributed in $[0, W]$, for $i = 1, 2, \cdots, M$,
and that each bucket is geometrically an $M$-dimensional cube with edge length 
$W$.  
For each $i = 1, 2, \cdots, M$, we define $x_i(-1)$ to be $\left\{\frac{f_i(\mathbf{q}) + b_i}{W}\right\} W$ (where $\left\{y\right\}$ denotes the fractional part of $y$),
$x_i(1)$ to be $W - x_i(-1)$, and $x_i(0)$ to be $0$.  Geometrically, $x_i(-1)$ and $x_i(1)$ are the distances between the epicenter 
$\mathbf{f(\mathbf{q})} + \mathbf{b}$ and the two ``faces'' of the epicenter cube (bucket) perpendicular to the $i^{th}$ dimension (axis).  
We denote as $\mathbf{x(\bm{\delta})}$ the distance vector $\langle x_1(\delta_1), x_2(\delta_2), \cdots, x_M(\delta_M)\rangle$.  

We now fix another arbitrary point $\mathbf{s}$. Suppose the Euclidean distance between $\mathbf{s}$ and $\mathbf{q}$ is $d_2$ (the subscript of which refers to $L_2$).  
Then since each $f_i$ is a Gaussian projection, $f_i(\mathbf{s}) - f_i(\mathbf{q})$ has a zero-mean Gaussian distribution with variance $d_2^2$.  
As a result, $Pr[h_i(\mathbf{s}) = h_i(\mathbf{q}) + 1]$, the probability with which $\mathbf{s}$ lands in the bucket whose $\delta_i$ (perturbation in the $i^{th}$ dimension) is $1$ 
is $\int_{x_i(1)}^{x_i(1)+W} \frac{1}{\sqrt{2\pi}d_2}e^{-x^2/2d_2^2}\,dx$.  It was shown in~\cite{lv2007multi} that this integral is roughly proportional to 
$e^{-x_i^2(1)/(2d_2^2)}$, the largest value that the integrand can attain in this interval. 
Similarly, $Pr[h_i(t) = h_i(q) - 1]$ (in which case $\delta_i = -1$) is roughly proportional to 
$e^{-x_i^2(-1)/(2d_2^2)}$. More generally, since hash functions $h_i$, $i=1, 2, \cdots, M$, are mutually independent,
the probability for $\mathbf{s}$ to land in a bucket with perturbation $\bm{\delta}$ is 
roughly proportional to $e^{-\|\mathbf{x(\bm{\delta})}\|_2^2/(2d_2^2)}$, where $\mathbf{x(\bm{\delta})}$ was defined above and 
$\|\mathbf{x(\bm{\delta})}\|_2^2 = \sum_{i = 1}^M x_i^2(\delta_i)$.

It will become clear shortly that $\|\mathbf{x(\bm{\delta})}\|_2^2$ is equal to the squared
Euclidean distance between the epicenter and the bucket with perturbation $\bm{\delta}$.
Since $\mathbf{s}$ is chosen arbitrarily (so $\mathbf{s}$ can be a nearest neighbor of $\mathbf{q}$), this approximation formula $e^{-\|\mathbf{x(\bm{\delta})}\|_2^2/(2d_2^2)}$ implies that 
the success probability of (finding $\mathbf{s}$ in) a bucket decreases when its squared (Euclidean) distance from the epicenter increases.
Hence the optimal probing sequence can instead be obtained by sorting these squared distances
in the increasing order.
These squared distances in turn have among them a simple additive structure that makes them very computationally efficient to compute and compare, as we explain in the following
toy example shown in Fig.\,\ref{fig:multi_probe}.
In this example, $M$ = 2, 
and the ``neighborhood'' of $\mathbf{h(q)}$
contains 8 equal-sized buckets.  Each bucket is geometrically a $W\times W$ rectangle (here $W=10$) and is 
represented by its perturbation vector.  For example, the bucket in the center with perturbation vector (0, 0) 
is the epicenter bucket.  In this example, the distances between the epicenter and the four ``faces'' 
are $x_1(1)=8.53$, $x_2(1)=4.62$, $x_1(-1)=1.47$, $x_2(-1)=5.38$ respectively.

We now explain the additive structure among these squared distance values, using this example.
We denote as $S^{(2)} = \{x_1(1)^2, x_2(1)^2, x_1(-1)^2, x_2(-1)^2\}$ the set of the 4 squared distances.
It is not hard to check that the squared distance from the epicenter to any of the 8 buckets in the ``neighborhood'' is a subset sum of the set $S^{(2)}$.
For example, the squared distance between the epicenter and the bucket $(1, 1)$ in Fig.\,\ref{fig:multi_probe}, denoted as $\|\mathbf{x}(1,1)\|_2^2$ is equal to $x_1(1)^2+x_2(1)^2 = 94.11$.   
Clearly, these subset sums are much easier to compute and compare than the corresponding success probabilities.

We now generalize this process to the case when $M$ is much larger.
There are 2M such distances at play, namely $x_i(-1)$ and $x_i(1)$ for $i = 1, 2, \cdots, M$.  We denote this set as $S$.  
Like in the toy example, let $S^{(2)}$ denote the set of these $2M$ squared distances.  
Again, the squared distance between the epicenter and each bucket in the ``neighborhood'' is a subset sum of $S^{(2)}$.
We denoted as $z_j$, $j = 1, 2, ..., 2M$, the $2M$ distances  
in $S$ sorted in the increasing order.  
Clearly the two smallest subset sums are $z_1^2$ and $z_2^2$ respectively.
For the third smallest subset sum, we need to compare $z_1^2 + z_2^2$ with $z_3^2$.  
As shown in~\cite{lv2007multi}, this search (for the next smallest) process can be implemented using the heap algorithm with such subset sums as keys.

\noindent
\textbf{The Third Refinement.} However, even to perform $O(T)$ heap operations is relatively computationally intensive.  
The third refinement is to precompute a universal (for all future queries) template from which a near-optimal probing sequence 
for any given query can be 
instantiated.  
Its idea is to perform the aforementioned search (for the next smallest subset sum) process under the 
idealized assumption that each $z_j^2$ (a random variable), $j = 1, 2, ..., 2M$, 
is equal to its expectation $E[z_j^2]$ (a constant);  
it was shown in~\cite{lv2007multi} that for $j, 1\leq j\leq M$, $E[z_j^2] = \frac{j(j+1)}{4(M+1)(M+2)}W^2$,
and for $j, M+1\leq j\leq 2M$, $E[z_j^2] = (1-\frac{2M+1-j}{M+1} + \frac{(2M+1-j)(2M+2-j)}{4(M+1)(M+2)})W^2$.
The resulting sorted list of subset sums 
is the universal template.
For example, when $M = 2$, 
the universal template is $[z_1^2, z_2^2, z_1^2 + z_2^2, z_3^2, z_1^2+z_3^2, z_4^2, z_2^2+z_4^2, z_3^2+z_4^2]$.

Now we explain how to instantiate a probing sequence from this template 
using the toy example shown in Fig.\,\ref{fig:multi_probe}.
Since $ x_1(-1) < x_2(1) < x_2(-1) < x_1(1)$, we know that $x_1(-1)$ is $z_1$, $x_2(1)$ is $z_2$, $x_2(-1)$ is $z_3$, $x_1(1)$ is $z_4$.  Hence $z_1^2$, the first element in the template, is instantiated to $x_1^2(-1)$, which corresponds to 
the bucket with perturbation $(-1, 0)$.
It is not hard to verify that the resulting probing sequence is $(-1, 0), (0, 1), (-1, 1), (0, -1), (-1, -1), (1, 0), (1, 1), (1, -1)$.
Using the third refinement, a near-optimal probing sequence can be computed two to three orders of magnitude faster than using only the 
first refinement.

\section{RW-LSH and Its Multi-Probe Extension}
In this section, we first describe random-walk LSH (RW-LSH), a new LSH scheme for ANNS-$L_1$.
Then we describe MP-RW-LSH, the multi-probe enhancement of RW-LSH.
Again, throughout this section, we focus on the operations in a single hash table.  

\subsection{The RW-LSH Scheme}\label{subsec:lsh-scheme}
To describe RW-LSH, we need to define what a random walk is.  Let $\tau^{(1)}$, $\tau^{(2)}, \cdots$ be a sequence of 
i.i.d. random variables, each of which takes value $1$ or $-1$ with equal probability $1/2$;  the value of each random variable is fixed 
once it is generated. The resulting (deterministic) sequence of values, denoted 
simply as $\tau$, is called a random walk. With a slight abuse of notation, we denote as $\tau(t)$ the position after $t$ steps along the random walk $\tau$ starting
at the origin;
that is, $\tau(t) \triangleq \tau^{(1)} + \cdots + \tau^{(t)}$.

It suffices to define a single raw hash value function $f$, since as explained earlier 
an RW-LSH function $h$ is derived from $f$ in the same way as in GP-LSH and CP-LSH:  $h(\cdot) = \left\lfloor \frac{f(\cdot) + b}{W} \right\rfloor$.
Suppose the dimension of the dataset $\mathcal{D}$ 
 is $m$.  
Then $f$ is a random walk projection parameterized by a vector of $m$ mutually independent random walks $\bm{\tau} = \langle \tau_1, \cdots, \tau_m \rangle$;
for the moment, we denote it as $f_{\bm{\tau}}$ to emphasize its dependence on $\bm{\tau}$.
Then given a data point $\mathbf{s} = \langle s_1, s_2, \cdots, s_m \rangle$, $f_{\bm{\tau}}(\mathbf{s})$ is defined as $\sum_{i=1}^m \tau_i(s_i)$.
We require that each $s_i$, $i = 1, 2,\cdots, m$, be a nonnegative even integer;  we will explain shortly why this assumption is not 
overly restrictive for real-world applications. 





Let $\mathbf{t}= \langle t_1, t_2, \cdots, t_m \rangle$ be another point in $\mathcal{D}$.
We denote as $d_1$ the value of the $L_1$ distance between $\mathbf{s}$ and $\mathbf{t}$, that is, $d_1 = \sum_{i=1}^m |s_i-t_i|$;
$d_1$ is a nonnegative even integer since each $s_i$ or $t_i$ is.
Then $f_{\bm{\tau}}(\mathbf{s}) - f_{\bm{\tau}}(\mathbf{t}) = \sum_{i=1}^m (\tau_i(s_i) - \tau_i(t_i))$ is a random walk of $\sum_{i=1}^m |s_i-t_i| = d_1$ steps, for the following 
reason:  For each $i$, $i = 1, 2,\cdots, m$, $\tau_i(s_i) - \tau_i(t_i)$ is an
$|s_i - t_i|$-step random walk along the sequence $\tau_i$, and these 
$m$ random walks are mutually independent since they are along different sequences. 
A subtle implication of this result is 
that for any two points $\mathbf{s}$ and $\mathbf{t}$, the random variable $f_{\bm{\tau}}(\mathbf{s}) - f_{\bm{\tau}}(\mathbf{t})$ is parameterized only by $\|\mathbf{s} - \mathbf{t}\|_1 = d_1$
(and not by $\mathbf{s}$ and $\mathbf{t}$).
For this reason, we can write $f_{\bm{\tau}}(\mathbf{s}) - f_{\bm{\tau}}(\mathbf{t})$ as $Y_{d_1}$, which has the following distribution:  $Pr[Y_{d_1} = l]$ is equal to $\binom{d_1}{(d_1+l)/2}\big{(}\frac{1}{2}\big{)}^{d_1}$ when $l$ is an even integer for 
$-d_1\le l \le d_1$, and is equal to $0$ otherwise. 

Similarly, we can show
that $h(\mathbf{s}) - h(\mathbf{t})$ is 
also parameterized only by $d_1$.  As a result, their collision probability 
$Pr[h(\mathbf{s}) = h(\mathbf{t})]$
is a function of $d_1$.
Hence, we can similarly (as we have defined $Y_{d_1}$) define $p(d_1)$ as the collision probability (when hashed by $h$) of any two points that are $d_1$ apart in $L_1$ distance.
It is not hard to verify that $p(d_1) = \sum_{l=-W}^{W}\left(1-\frac{\lvert l \rvert}{W}\right)Pr[Y_{d_1} = l]$, which is 
a convolution between the uniform distribution (the first term in the summand)
and the random-walk distribution (the second term).
 It is not hard to prove that,
 when $W$ is a positive even integer, the collision probability $p(d_1)$ decreases
monotonically when $d_1$ takes on only nonnegative even integer values (which $d_1$ indeed does under our assumptions)
that is, $p(0) > p(2) > p(4) > \cdots$. The proof
can be found in Sect\,\ref{subsec:proof}.
With this monotonicity property, RW-LSH meets the standard requirement (needed to ensure that $p_2 < p_1$) to qualify as an LSH family.

\subsection{Discussions on RW-LSH}\label{subsec:rw_lsh_issue}
Although we restrict the domain of each coordinate value $s_i$ 
to nonnegative even integers, the RW-LSH scheme 
can be extended to work without this restriction as follows.
First, for each dimension $i$, we can increment (shift) the $i^{th}$ coordinate of every point in $\mathcal{D}$ by a large enough positive constant $a_i$
 so that these $i^{th}$ coordinates all become nonnegative.
Second, we can multiply (scale) every vector by a large enough integer number $c$ and then round each resulting scalar to the nearest even integer.  
It is clear that both the shift and the scaling operations preserve the ranked order among the $L_1$ distance values. 
Although rounding can cause changes to this ranked order, the percentage of such changes can be made extremely small, by increasing the $c$ value, 
so that with overwhelming probability, an ANNS query over the original dataset has the same correct answer  
as that over the rounded scaled shifted dataset.


We now discuss an implementation issue of RW-LSH. As usual, each random walk sequence $\tau_i$ (for implementing a function $f_{\bm{\tau}}$) is implemented as a pseudorandom bit 
sequence wherein bit $0$ is interpreted as $-1$. It certainly does not make sense to regenerate these $m$ pseudorandom sequences when computing $f_{{\bm\tau}}(\mathbf{q})$ for each query point 
$\mathbf{q}$. Hence we precompute and store each $\tau_i(t)$ for $t = 2, 4, 6, \cdots, U_i$ where $U_i$ is the maximum possible (even) 
value for the $i^{th}$ coordinate of a data point.  Let the {\it universe} $U$ be the maximum value
among $U_1,U_2,\cdots,U_m$.  For each hash table (with $M$ hash functions), we need a maximum of $mUM$ bytes for storing the precomputed table (one table entry costs 2 bytes for each even $t$ value).  
For most real-world datasets, this storage cost is small (typically more than an order of magnitude smaller) relative to the size of each hash table, 
especially when the dataset is large enough to pose a scalability challenge, since this cost is fixed in the sense
it is independent of the size of the dataset $\mathcal{D}$.  
For example, this cost increases the index size by only $0.4\%$ for a 50 million-point dataset used in our experiments described in 
Sect.\,\ref{sec:eva}.  Hence, we do not include this storage cost in the index sizes reported in Sect.\,\ref{subsec:comp_lsh}
since it does not alter the scalability narrative.

\subsection{Multi-Probe Extension}\label{subsec:multi_probe_extension}
From this point on, we drop the subscript $\bm{\tau}$ from $f_{\bm{\tau}}$. 
The multi-probe extension of RW-LSH (to MP-RW-LSH) is straightforward: It is identical to that of GP-LSH.
This ``porting'' is possible for the following reason.  
Recall that in both RW-LSH and GP-LSH, an LSH function $h$ is defined as
$h(\cdot)=\left\lfloor \frac{f(\cdot)+ b}{W} \right\rfloor\ $. 
They differ only in (the choice of) the raw hash value function $f$.
Recall that the following property of a Gaussian projection $f$ 
is the sufficient condition for all three multi-probe refinements to work for GP-LSH: For any $\mathbf{s}$ and $\mathbf{t}$, $f(\mathbf{s}) - f(\mathbf{t})$
has a zero-mean Gaussian distribution (with variance $d_2^2 = \|\mathbf{s} - \mathbf{t}\|_2^2$).  
However, this zero-mean Gaussian distribution (with variance $d_1 = \|\mathbf{s} - \mathbf{t}\|_1$) property continues to hold approximately when $f$ is instead a 
random walk projection, 
when $d_1$ is large.

Readers naturally would ask ``What if $d_1$ is small?''  To answer this question, let $\mathbf{s}$ be a point in $\mathcal{D}$ and $\mathbf{q}$ be a query point and let $d_1$ be
their $L_1$ distance.   
Then $f(\mathbf{s})+b$ is just $d_1$ random walk steps away from the epicenter $f(\mathbf{q})+b$.  When $d_1$ is small, $f(\mathbf{s})+b$  is known to be 
slightly statistically closer to $f(\mathbf{q})+b$ than if $f(\mathbf{s}) - f(\mathbf{q})$ is precisely Gaussian. 
As a result, $h(\mathbf{s})$ tends to fall into the epicenter bucket $h(\mathbf{t})$ with a higher probability. Hence the net effect of this approximation error 
is to allow the epicenter bucket to ``steal'' small amounts of success probability (of finding a point like $\mathbf{s}$) from the
``neighboring'' buckets. This does not negatively impact the efficacy of multi-probe since the epicenter bucket is to be probed anyway. 

\section{Why CP-LSH Is Not Suitable for Multi-Probe?}\label{subsec:cauchy}
We have found that the quality value $\rho$ of RW-LSH is slightly larger (worse) than that of CP-LSH.   
Recall that for any standard LSH scheme, its query time complexity is $O(n^\rho)$ and the space complexity in terms of number of hash tables is also $O(n^\rho)$.
Hence RW-LSH would perform slightly worse than CP-LSH in terms of both complexities. 
We will show that MP-RW-LSH can successfully reduce the number of hash tables to almost a
constant (typically between 6 and 8), 
so its quality $\rho$ will not affect its space complexity anymore. 
However, its time complexity remains $O(n^\rho)$, since $O(n^\rho)$ buckets still have 
to be probed except that these buckets are now spread over 6 to 8 (instead of $O(n^\rho)$) hash tables.  
This, combined with a slightly larger $\rho$ value for RW-LSH, 
explains why the query time of MP-RW-LSH is slightly higher that of CP-LSH shown in Sect.\,\ref{subsec:comp_lsh}. 

\begin{wraptable}{R}{0.59\textwidth}
\centering
 \caption{$\mathcal{P}_T(d_1)$ w/ optimal probing sequences.
 } \label{tab:rw_cauchy_probability}
 \resizebox{0.59\textwidth}{!}{\begin{tabular}{@{}|l|r r r|r r r|@{}}
\hline
\multirow{2}{*}{$d_1$} & \multicolumn{3}{c|}{MP-RW-LSH} & \multicolumn{3}{c|}{MP-CP-LSH} \\ \cline{2-7}
  & T=30 & T=60 & T=100 & T=30 & T=60 & T=100\\ 
 \hline
6   & 0.50  & 0.63 & 0.72  & 0.0405 & 0.0568 & 0.0716 \\
8   & 0.36 & 0.48 & 0.57 & 0.0137 & 0.0203 & 0.0268 \\
12   & 0.19 & 0.27 & 0.34  & 0.0018 & 0.0030 & 0.0043 \\ 
16   & 0.10  & 0.15 & 0.20 & 0.0003 & 0.0005 & 0.0008 \\ 
  \hline
 \end{tabular}}
\end{wraptable}



We now explain why, despite that RW-LSH has worse quality $\rho$ than CP-LSH, RW-LSH is much better suited for multi-probe extension than CP-LSH.
We do so by comparing the success probabilities, defined precisely next, of their respective multi-probe extensions MP-RW-LSH and MP-CP-LSH.
Let $\mathbf{q}$ be a query point, from which the values of the   
epicenter ($\mathbf{f(q)+b}$), the epicenter bucket ($\mathbf{h(q)}$), the perturbation vector $\bm{\delta}$ and the distance vector $\mathbf{x(\bm{\delta})}$ are derived. 
Let $\mathbf{s}$ be a nearest neighbor of $\mathbf{q}$ (in $\mathcal{D}$) and $d_1$ be their $L_1$ distance.  
We denote as $\mathcal{P}_T(d_1)$ the total success probability of (finding $\mathbf{s}$ in) the top-(T+1) buckets along a probing sequence.
We compare $\mathcal{P}_T(d_1)$ values under MP-RW-LSH and MP-CP-LSH with the respective optimal probing sequences.
For a fair comparison, $M$ is set to a typical value of $10$ in both baselines RW-LSH and CP-LSH, and $W$ is set to $8$ in RW-LSH and $20$ in CP-LSH to 
achieve an optimal or near-optimal $\rho$ value for $r_1 = 6$ (near radius) and $r_2 = 12$ (far radius).

\begin{wraptable}{R}{0.32\textwidth}
\centering
 \caption{ $\mathcal{P}_T(d_1)$ w/ template-generated probing sequence.} \label{tab:rw_temp_probability}
 \resizebox{0.32\textwidth}{!}{\begin{tabular}{@{}|l|r r r|@{}}
\hline
\multirow{2}{*}{$d_1$} &\multicolumn{3}{c|}{MP-RW-LSH}\\
\cline{2-4}
 & T=30 & T=60 & T=100 \\ 
 \hline
6   & 0.46  & 0.58 & 0.67 \\
8   & 0.33 & 0.43 & 0.52 \\
12   & 0.17  & 0.24 & 0.31 \\ 
16   & 0.09  & 0.14 & 0.19 \\  
  \hline
 \end{tabular}}
\end{wraptable}

The comparison results are shown in Table~\ref{tab:rw_cauchy_probability}. 
Each number in Table~\ref{tab:rw_cauchy_probability} is the average of 1,000 simulation runs.  For each variant, we calculate and demonstrate in Table~\ref{tab:rw_cauchy_probability} the 
$\mathcal{P}_T(d_1)$ values for the following 12 value combinations of $d_1$ and $T$: $d_1 = 6, 8, 12, 16$ and $T = 30, 60, 100$.  
Table~\ref{tab:rw_cauchy_probability} clearly shows, for the same $T$ and $d_1$, the $\mathcal{P}_T(d_1)$ under MP-CP-LSH
is one to two orders of magnitude smaller than that under MP-RW-LSH;  this ``top-light'' behavior of MP-CP-LSH is expected since the
Cauchy distribution underlying CP-LSH is heavy-tailed~\cite{resnick2007heavy}.
As a result, MP-CP-LSH would need a much larger number of hash tables to achieve the same query accuracy (success probability) as MP-RW-LSH.
For example, when $T = 100$ and $d_1 = 8$, MP-RW-LSH needs to use only 6 hash tables to achieve a success probability of $1-(1-0.57)^6 = 0.99$, whereas MP-CP-LSH 
needs to use 186 hash tables to do the same.

In our evaluations next, for MP-RW-LSH, we use a precomputed template (the third refinement described in Sect.\,\ref{subsec:mplsh}) 
to generate a near-optimal probing sequence given any 
query $\textbf{q}$, so that the query time can be minimized.  This optimization would sacrifice the query accuracy only slightly, as we have verified using simulations.  
In this simulation study, we use the same experimental setup and parameter settings, except that the probing sequences
are generated by the precomputed template.  The results, shown in Table~\ref{tab:rw_temp_probability}, demonstrate that using the template-generated probing sequences
reduces the success probability values shown in Table~\ref{tab:rw_cauchy_probability} by only
$5\%$ to $10\%$.

\section{Evaluation}\label{sec:eva}
In this section, we evaluate the ANNS-$L_1$ query performance of MP-RW-LSH 
against those of the following three LSH schemes:
CP-LSH, RW-LSH (its own baseline), and SRS~\cite{Sun16_thesis}.
All four algorithms are implemented and optimized for in-memory operations, and are hence evaluated as such.
A few other LSH-based ANNS-$L_1$ algorithms, such as QALSH~\cite{huang2017query}, are not compared here, 
since they are implemented and optimized for external-memory operations.
CP-LSH (in terms of query efficiency) and SRS (in terms of scalability) are two state-of-the-art LSH solutions for ANNS-$L_1$.
Our evaluations show conclusively that although its baseline RW-LSH is ``mediocre'' compared to CP-LSH and SRS,
MP-RW-LSH achieves a much better tradeoff between the query efficiency and scalability than both CP-LSH and SRS.

\subsection{Experiment Settings}\label{subsec:exp_set}
\noindent
\textbf{Evaluation Datasets.}
\begin{wraptable}{R}{0.61\textwidth}
\caption{Datasets summary.}\label{tab:dataset-stats}
\centering
\resizebox{0.61\textwidth}{!}{\begin{tabular}{@{}ll cccccc@{}}
\specialrule{.1em}{.05em}{.05em} 
 \multicolumn{2}{c}{Dataset} & $n$ & $m$ & $n_q$ & $U$ & Type\\ 
\specialrule{.1em}{.05em}{.05em} 
  \multirow{3}{*}{Small} &Audio~\cite{audio} &53.3K & \num{192} & 200 & 200K &Audio\\
&MNIST~\cite{mnist} & 69.0K & \num{784} & 200 &2K &Image\\ 
&Trevi~\cite{Trevi} & 99.9K & \num{4096} & 200 &510 &Image\\ 
\specialrule{.1em}{.05em}{.05em}
 \multirow{2}{*}{Medium} &GIST~\cite{sift-gist} & 1.0M & \num{960} & 1K &3K &Image\\ 
&GloVe~\cite{glove} & 1.2M & \num{100} & \num{200} &25K &Text\\ 
\specialrule{.1em}{.05em}{.05em} 
  \multirow{2}{*}{Large} &Deep10M~\cite{deep} & 10.0M & \num{96} & 10K &3K &Image\\
&SIFT50M~\cite{sift-gist} & 50.0M & \num{128} & 10K &510 &Image\\ 
\specialrule{.1em}{.05em}{.05em} 
\end{tabular}}
\end{wraptable}
We use seven widely used publicly available datasets of diverse dimensions, sizes (number of points), and types.
The SIFT50M dataset contains 50 million points sampled uniformly at random from the 1 billion points contained in SIFT1B~\cite{sift-gist}. We cannot use SIFT1B instead since 
the resulting index structures of CP-LSH and RW-LSH would not fit into the main memory. 
We normalize (scale and round as described in Sect.\,\ref{subsec:rw_lsh_issue}) the coordinates of all data points to nonnegative even integers in all seven datasets. For each of 
the seven nominalized datasets, Table~\ref{tab:dataset-stats} shows its size $n$, its dimension $m$, the number of queries $n_q$ processed on it, 
its universe $U$ (defined in Sect.\,\ref{subsec:rw_lsh_issue})
 and its type.
We drop the word ``normalized'' in the sequel with 
the understanding that all datasets we refer to by names have been normalized.

\noindent\textbf{Performance Metrics.} 
We evaluate the performances of these four algorithms in three aspects:  scalability, query efficiency, and query accuracy.  
To measure scalability (how well an algorithm can scale to very large datasets), we use the {\it index size} (excluding the size of the original dataset).
For each query, each algorithm being evaluated needs to find $k = 50$ nearest neighbors in $L_1$ distance.  
To measure query efficiency, we use {\it query time}.
To measure query accuracy, we use {\it recall} and {\it overall ratio}, 
defined as follows.  For a query point $\mathbf{q}$, let the query result be $R = \{\mathbf{o_1}, \mathbf{o_2}, \cdots, \mathbf{o_k}\}$ with its elements sorted in 
the increasing order of their $L_1$ distances to $\mathbf{q}$, and $R^*=\{\mathbf{{o_1}^*}, \mathbf{{o_2}^*}, \cdots, \mathbf{{o_k}^*}\}$ be the actual $k$ nearest neighbors similarly sorted. 
The overall ratio and recall are defined as $\frac{1}{k} \sum_{i=1}^k ||\mathbf{q}- \mathbf{o_i}||_1 / ||\mathbf{q}-\mathbf{{o_i}^*}||_1$ and $|R \cap R^{*}| / |R|$, respectively. 
Each query time, recall, or overall ratio value presented in Table~\ref{tab:query} and Fig.\,\ref{fig:table_recall} is the average over many queries. 

\noindent
\textbf{Implementation Details.}
We implement RW-LSH functions, CP-LSH functions, and the multi-probe framework with template-generated probing sequence 
in C++. 
For indexing and querying in LSH, we use an efficient open-source C++ LSH implementation called FALCONN\footnote{\url{https://github.com/FALCONN-LIB/FALCONN}}. 
For SRS, we use the C++ source code provided by its authors. We compile all C++ source codes using g++ 7.5 with -O3.  
All experiments are done on a workstation running Ubuntu 18.04 with 
Intel(R) Core(TM) i7-9800X \SI{3.8}{GHz} CPU, \SI{128}{GB} DRAM
and \SI{4}{TB} hard disk drive (HDD).

\subsection{Benchmark Algorithms and Parameter Settings}\label{subsec:ann-comp-algorithms}
We first briefly describe SRS~\cite{Sun16_thesis}, the only benchmark algorithm that has not been introduced before. 
In SRS, at the indexing stage, 
each point $\mathbf{s} \in \mathcal{D}$ is mapped to an $M$-dimensional vector $\mathbf{f(s)} = \langle f_1(\mathbf{s}), f_2(\mathbf{s}), \cdots, f_M(\mathbf{s})\rangle$,
where each $f_i$ is a Cauchy projection like that in CP-LSH and $M$ is typically between 6 and 10. 
Then given a query $\mathbf{q}$, SRS searches in the ``projection image'' $\mathbf{f}(\mathcal{D})$ for $t$ exact nearest
neighbors ($t$-NN) of $\mathbf{f(q)}$, where $\mathbf{f}(\mathcal{D})\triangleq\{ \mathbf{f(t)} \mid \mathbf{t} \in \mathcal{D}\}$.   This $t$-NN search can be computed very
efficiently by organizing $\mathbf{f}(\mathcal{D})$, a low-dimensional point set, as a cover tree.

Now for each algorithm, we describe how we tune its parameters for the best query performance. 
In RW-LSH, MP-RW-LSH, and CP-LSH, we have three parameters to tune:  $M$ (the dimension of an LSH function vector), 
$W$ (the bucket ``width''), and $L$ (the number of hash tables). In SRS, we have two parameters to tune: $M$ and $t$ (defined above
in ``$t$-NN search''). There is no $L$ in SRS, since it uses a cover tree instead of hash tables as the index structure.

\noindent
{\bf RW-LSH and MP-RW-LSH.}  
For RW-LSH, we find that the following value combinations of $(M, W)$ strike the best tradeoffs between query accuracy and query efficiency for the seven datasets 
listed in Table~\ref{tab:dataset-stats} from top to bottom respectively:
$(12, 3144), (12, 930), (16, 1728), (8, 452), (16, 1104), (17, 424), (14, 224)$. The same value combinations are used for MP-RW-LSH.
MP-RW-LSH has an additional parameter to tune:  $T$ (number of additional buckets to be probed in each hash table). 
We find that $T=100$ strikes near-optimal tradeoffs between query time and query accuracy for all seven datasets. For both RW-LSH and 
MP-RW-LSH, we adjust $L$ to achieve a certain level of query accuracy for each dataset. 

\noindent
{\bf CP-LSH.} We use the following near-optimal parameter settings for the seven datasets in the same order as above: $(M, W) = (7, 4401336),(8, 384416),
(7, 561426),\\(6, 153732),(8, 303312),(6, 34014),(8, 17336)$.
For each
dataset, we adjust $L$ to achieve a similar query accuracy as achieved by RW-LSH and MP-RW-LSH.



\noindent
\textbf{SRS.}
It was suggested by authors of SRS that $M$ should range from 6 to 10~\cite{Sun16_thesis}.
For all seven datasets, we find that $M=10$ strikes roughly the best tradeoffs between query accuracy and query efficiency. 
As suggested by authors of SRS~\cite{Sun16_thesis}, 
we adjust parameter $t$ to reach the same level of 
query accuracy as achieved by the other three algorithms for each dataset.  

\subsection{Comparison between MP-RW-LSH, CP-LSH, and RW-LSH}\label{subsec:comp_lsh}
In this section, we compare MP-RW-LSH, CP-LSH, and RW-LSH in terms of scalability and query efficiency.
We compare MP-RW-LSH with SRS separately because, unlike the other three, SRS uses a cover tree instead of hash tables as the index structure. 
In Table~\ref{tab:query}, we report the query times and the index sizes needed by all four algorithms for achieving similar query accuracies (if possible) on each dataset.

\begin{figure}[htb]
\centering
\begin{subfigure}[t]{0.5\textwidth}
    \centering
    \includegraphics[width=\textwidth]{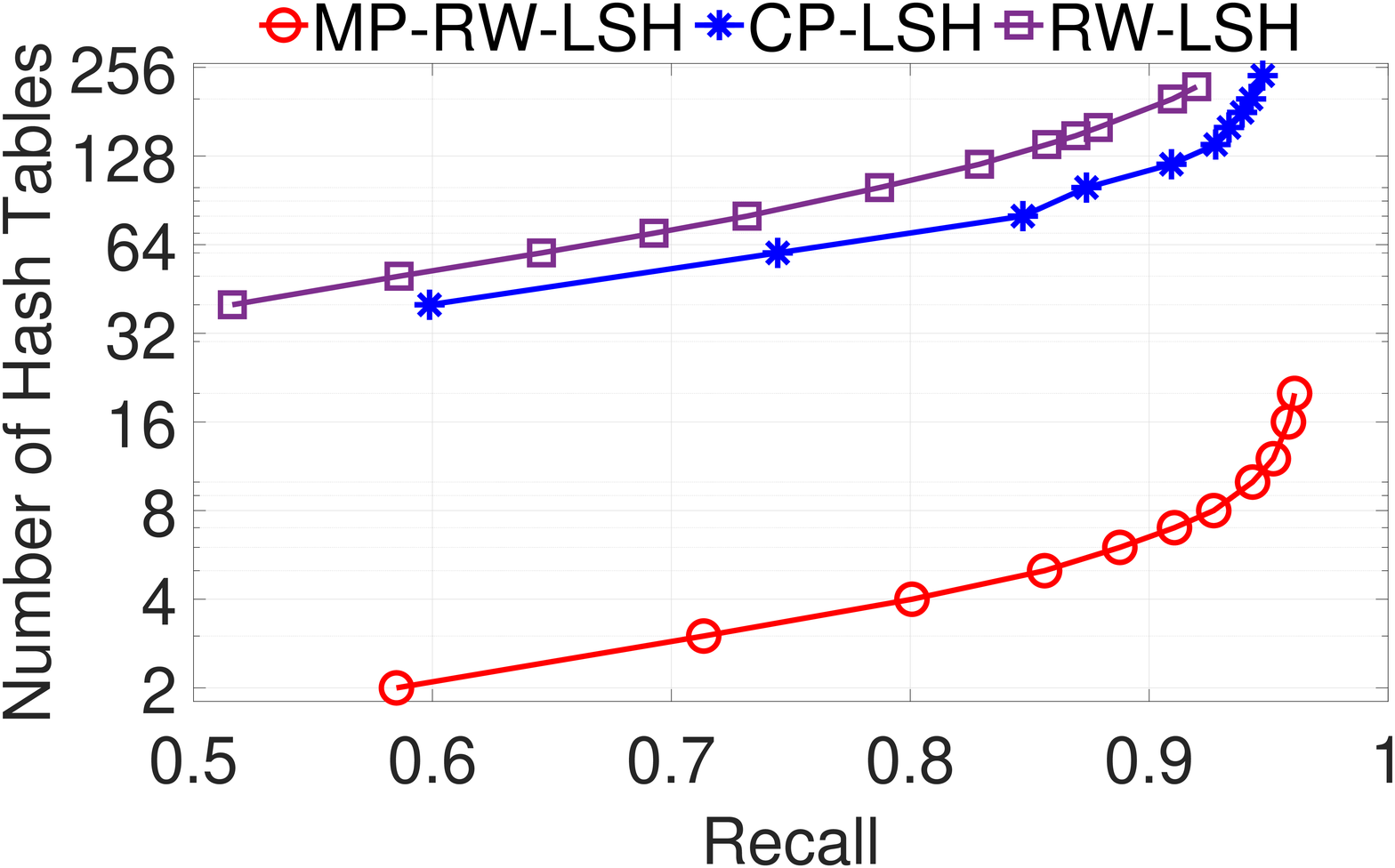}
    \caption{GIST}\label{fig:table_recall_gist}
\end{subfigure}%
\begin{subfigure}[t]{0.5\textwidth}
    \centering
    \includegraphics[width=\textwidth]{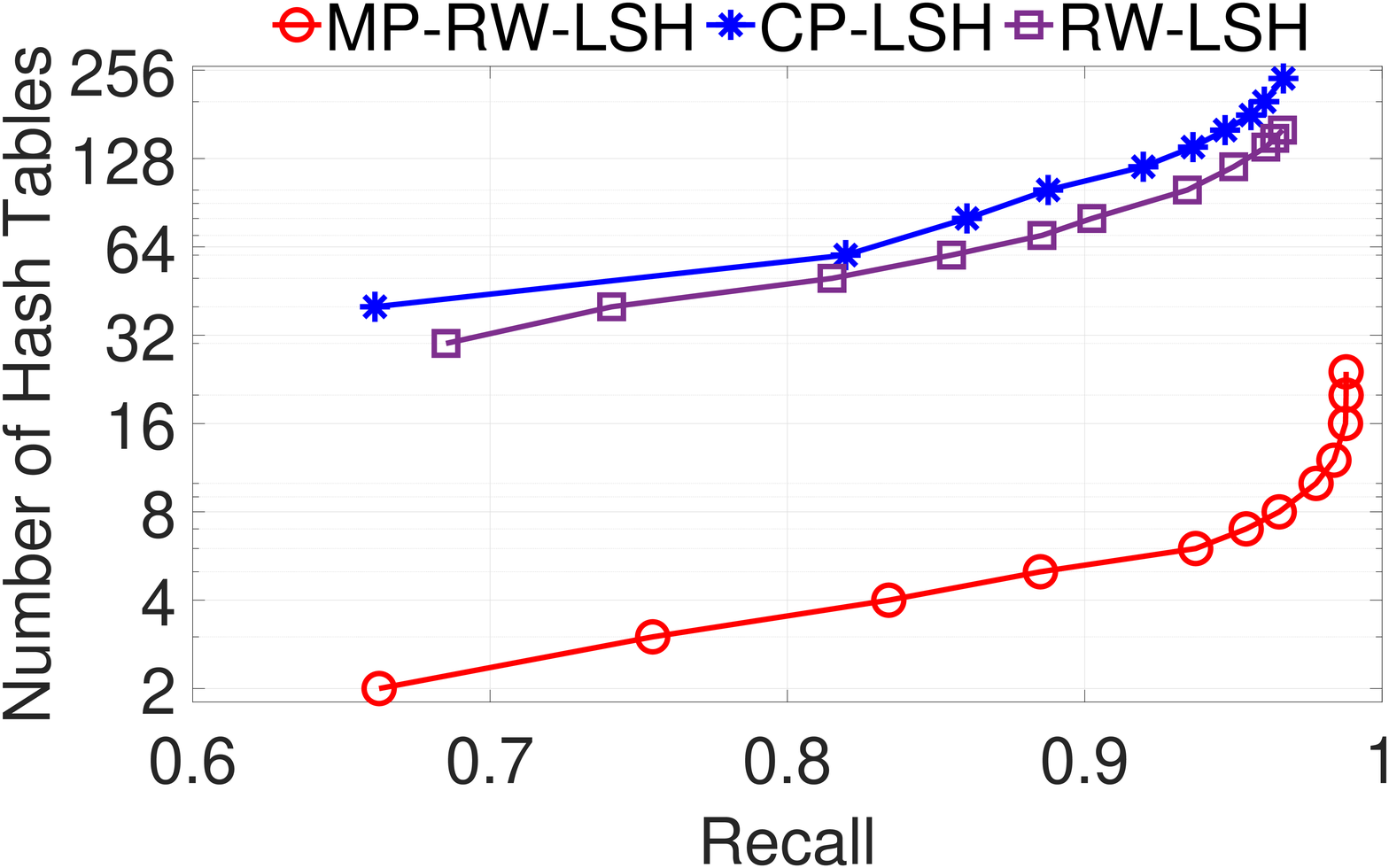}
    \caption{GloVe} \label{fig:table_recall_glove}
\end{subfigure}
\caption{Number of hash tables vs. recall.}\label{fig:table_recall}
\end{figure}

\noindent
{\bf Scalability.} 
Table~\ref{tab:query} clearly shows that MP-RW-LSH has much better scalability than both CP-LSH and RW-LSH.   
On all seven datasets, the index sizes of MP-RW-LSH are 14.8 to 53.3 and 15.0 to 27.5 times smaller than those of CP-LSH and RW-LSH, respectively. 
The numbers of hash tables used in the MP-RW-LSH are also smaller than those used in CP-LSH and RW-LSH by the same ratios, since 
this number is proportional to the index size in all three algorithms for the same dataset.
Fig.\,\ref{fig:table_recall} shows the tradeoffs between
recall values achieved and the numbers of hash tables used by the three algorithms on two medium datasets
GIST and GloVe.  Fig.\,\ref{fig:table_recall_gist} shows that for achieving the same recall value, CP-LSH and RW-LSH need to use roughly 18.2 to 20.1 and roughly 24.8 to 27.5 times more hash tables than MP-RW-LSH on GIST, respectively.
Fig.\,\ref{fig:table_recall_glove} shows that for achieving the same recall value, CP-LSH and RW-LSH need to use roughly 20.1 to 29.2 and roughly 13.9 to 19.4 times more hash tables than MP-RW-LSH on GloVe, respectively.
In fact, MP-RW-LSH can scale to the one-billion-point dataset SIFT1B (without sampling)~\cite{sift-gist} with an index size of roughly \SI{24}{GB}, whereas neither CP-LSH nor RW-LSH can 
(using the \SI{128}{GB} memory the computer has)
while achieving the same query accuracy as MP-RW-LSH. 

\noindent
{\bf Query Efficiency.}  
As shown in Table~\ref{tab:query}, for achieving similar (or better) query accuracies,
MP-RW-LSH has shorter query times on all the three small datasets and similar or slightly longer query times on all medium and large datasets than its baseline RW-LSH.
Table~\ref{tab:query} also shows that CP-LSH has between 1.3 to 2.2 times shorter query times than MP-RW-LSH on the seven datasets, the reason for which was
explained in the first paragraph in Sect.\,\ref{subsec:cauchy}. Overall, it is fair to say that MP-RW-LSH achieves a much better tradeoff between scalability and query efficiency than CP-LSH.
\begin{table}[!htb]
\caption{Experiment results overview.}\label{tab:query}
\centering
 \begin{tabular} {@{}|c|c|c|c|c|@{\hspace{8pt}}c@{\hspace{8pt}}|@{}}
\specialrule{.1em}{.05em}{.05em} 
 \multicolumn{1}{|c}{} && \textbf{MP-RW-LSH} & \textbf{CP-LSH} & \textbf{RW-LSH} &  \textbf{SRS}  \\ \specialrule{.1em}{.05em}{.05em} 
\multirow{4}{*}{Audio}  & Query Time (ms)  & 4.2 & 2.4  & 13.5 & 18.0  \\  \cline{2-6}
& Recall & 0.9307 & 0.9298 & 0.8445 & 0.9140  \\  \cline{2-6}
& Overall Ratio  & 1.0032 & 1.0033 & 1.0091 & 1.0050 \\ \cline{2-6}
& Index Size (MB) & 65.6 & 968.0 & 984.4 & 3.0  \\ \specialrule{.1em}{.05em}{.05em} 
\multirow{4}{*}{MNIST} & Query Time (ms)  & 11.6 & 5.6 & 37.1 & 47.7 \\  \cline{2-6}
& Recall & 0.9333 & 0.9309 & 0.9221 & 0.9314  \\  \cline{2-6}
& Overall Ratio  & 1.0046 & 1.0048 & 1.0056 & 1.0070  \\ \cline{2-6}
& Index Size (MB) & 66.1 & 2644.2 & 1487.4 & 3.8  \\ \specialrule{.1em}{.05em}{.05em} 
\multirow{4}{*}{Trevi}  & Query Time (ms)  & 60.5 & 40.4 & 147.1 & 86.8  \\  \cline{2-6}
& Recall & 0.9187 & 0.9162 & 0.9055 & 0.9168  \\  \cline{2-6}
& Overall Ratio  & 1.0035 & 1.0038 & 1.0036 & 1.0042  \\  \cline{2-6}
&Index Size (MB)  & 50.3 & 2681.9  & 1005.7& 5.7  \\ \specialrule{.1em}{.05em}{.05em} 
\multirow{4}{*}{GIST}   & Query Time (ms)  & 354.3 & 247.0 & 364.1 & 1045.8  \\  \cline{2-6}
& Recall & 0.9630 & 0.9576 & 0.9557 & 0.9602  \\  \cline{2-6}
& Overall Ratio  & 1.0009 & 1.0010 & 1.0010 & 1.0010  \\  \cline{2-6}
& Index Size (MB)  & 94.5 & 1417.8 & 2599.2  & 52.6  \\ \specialrule{.1em}{.05em}{.05em} 
\multirow{4}{*}{GloVe}  & Query Time (ms)  & 152.0 & 119.2 & 143.6 & 557.1  \\  \cline{2-6}
& Recall & 0.9766 & 0.9753 & 0.9751	 & 0.9648  \\  \cline{2-6}
& Overall Ratio  & 1.0006 & 1.0007 & 1.0005 & 1.0010  \\  \cline{2-6}
& Index Size (MB)  & 100.4  & 3764.7  & 1882.4  & 52.6  \\ \specialrule{.1em}{.05em}{.05em} 
\multirow{4}{*}{Deep10M}  & Query Time (ms)  & 1045.0 & 560.8 & 825.6 & 5338.8 \\ \cline{2-6}
& Recall & 0.9756 & 0.9758 & 0.9737 & 0.9565  \\  \cline{2-6}
& Overall Ratio  & 1.0008 & 1.0007 & 1.0008 & 1.0017 \\ \cline{2-6}
& Index Size (MB)  & 323.0 & 6922.0 & 5537.6 & 525.2  \\ \specialrule{.1em}{.05em}{.05em}  
\multirow{4}{*}{SIFT50M}  & Query Time (ms)  & 5475.7 & 2445.0 & 3615.4 & 28302.9  \\  \cline{2-6}
& Recall & 0.9807 & 0.9809 & 0.9668 & 0.9595  \\  \cline{2-6}
& Overall Ratio & 1.0006 & 1.0005 & 1.0011 & 1.0017  \\ \cline{2-6}
& Index Size (MB) & 1192.4 & 19873.5 & 17886.1 & 2656.8  \\ \specialrule{.1em}{.05em}{.05em} 
 \end{tabular}
\end{table}


\subsection{Comparison between MP-RW-LSH and SRS}
\noindent
{\bf Scalability.} 
As shown in Table\,\ref{tab:query}, MP-RW-LSH has smaller index sizes on the two large datasets, but has larger index sizes on the five small and medium datasets, than SRS.  
However, MP-RW-LSH is actually fundamentally more scalable than SRS, and its larger index sizes on the five small and medium datasets is due to the following reason. 
For all seven datasets, we fix the number of hash
buckets in each hash table to $2^{21} \approx$ 2.1 million in MP-RW-LSH, CP-LSH and RW-LSH, because we do not want to throw in another tunable parameter to
``muddle the water'' of optimizing the parameters of these three algorithms for fair comparisons. Hence there is a fixed (i.e., not growing with $n$) cost of \SI{8}{MB} per hash table 
for storing the array of 2.1 million head cells each of which contains a 4-byte-long pointer to the data points hashed into the corresponding bucket.  
Excluding this fixed cost, MP-RW-LSH indeed has smaller index sizes than SRS on the five small and medium datasets.  
For example, on Audio for which MP-RW-LSH uses 8 hash tables, this fixed cost is \SI{64}{MB}; excluding this fixed cost, 
the index size of MP-RW-LSH becomes $65.6 - 64 =$ \SI{1.6}{MB}, which is smaller than that of SRS at \SI{3.0}{MB}.   

\noindent
{\bf Query Efficiency.}
As shown in Table~\ref{tab:query}, MP-RW-LSH has much shorter query times than SRS for achieving similar 
query accuracies, especially on large datasets.
Therefore, it is fair to say that overall MP-RW-LSH achieves a much better tradeoff between scalability and query efficiency than SRS.

\section{Related Work}
There are many ANNS algorithms based on different techniques.Here, we only list the LSH algorithms focusing on ANNS-$L_1$ in the interest of space.
To avoid the large indexing size of plain-vanilla LSH schemes, Sun {\it et al.}~\cite{Sun16_thesis} introduce a projection-based method named SRS for ANN-$L_1$.
Similar to CP-LSH, SRS also use Cauchy projection to project a point $s$ as a $M$-dimensional (typically 6 to 10) raw hash vector $\mathbf{f(s)} = \langle f_1(\mathbf{s}),f_2(\mathbf{s}),\cdots,f_M(\mathbf{s})\rangle$. 
This projection maps the original dataset $\mathcal{D}$ that lies in a high-dimensional space  to $\mathbf{f}(\mathcal{D})$ that lies in a low-dimensional space. 
Since Cauchy projections is statistically closeness-preserving 
in the sense that if the point $\mathbf{s}$ is among the closest points to the query point
$\mathbf{q}$ in $L_1$ distance, then $\mathbf{f(s)}$ should statistically be among the closest to $\mathbf{f(q)}$ in $L_1$ distance. 
An ANN query over the high-dimensional dataset $\mathcal{D}$ is thus converted to a t exact nearest
neighbors ($t$-NN) search 
over the low-dimensional dataset
$\mathbf{f}(\mathcal{D})$. The latter $t$-NN search can be computed very
efficiently through a cover tree due to the dimension of $\mathbf{f}(\mathcal{D})$ is typically 6 to 10. However,  $t$ can be very large to achieve a high query quality, since the statistical closeness preservation is not very accurate
which makes the nearest neighbor $\mathbf{s}$ of $\mathbf{q}$ can be far away in low dimensional space.

In addition to SRS described above, many recent works, for example QALSH~\cite{huang2017query}, have adapted LSH for external-memory operations, which can solve ANN search in $L_1$ distance.
These LSH algorithms typically need a large number of hash tables (e.g. 184~\cite{huang2017query}), so their index sizes are too large to fit in memory. Therefore, they have to use disk-resident data structures, which results in long query time.

\section{Conclusion}
In this paper, we propose MP-RW-LSH, the first and so far only multi-probe LSH solution for ANNS-$L_1$ distance. 
We show that to achieve a similar query accuracy and comparable query efficiency, MP-RW-LSH has much smaller index structures and much less hash tables than CP-LSH,
which allows it to scale to much larger datasets. 
We also show that MP-RW-LSH has smaller index sizes for large datasets and achieves a much better query efficiency than SRS for achieving similar query accuracies.

\bibliographystyle{splncs04}
\bibliography{bib/NTGfull,bib/lsh,bib/applications,bib/theory,bib/dim_reduction,bib/dataset}

\section{Appendix}\label{sec:appendix}
\subsection{Proof of monotonically decreasing property}\label{subsec:proof}
In the following proof, we drop the subscript $1$ from $d_1$ in both places they appear in:  $p(d_1)$ and $Y_{d_1}$.
\begin{proof}
It suffices to prove that, for any nonnegative even integer $d$, we have $p(d)>p(d+2)$.  We have
\begin{eqnarray} \nonumber
\lefteqn{p(d) = \sum_{\ell=-W}^{W}\left(1-\frac{|\ell|}{W}\right)\Pr[Y_d = \ell] } \\\nonumber
&=&  \sum_{\ell=0}^{W}\left(1-\frac{\ell}{W}\right)\Pr[|Y_d| = \ell] \\ \nonumber
&=& \sum_{\ell=0}^{W-1} \sum_{t=0}^{W-\ell-1} \frac{1}{W} \Pr[|Y_d| = \ell]  \\ \nonumber
&=& \sum_{t=0}^{W-1} \sum_{\ell=0}^{W-t-1} \frac{1}{W} \Pr[|Y_d| = \ell]  \\ \nonumber
&=& \frac{1}{W}  \sum_{t=0}^{W-1} \Pr[|Y_d| \leq W-t-1] \\ \label{eq:coll_prob_prove_part1}
&=& \frac{1}{W}  \sum_{t=0}^{W-1} \Pr[|Y_d| \leq t] \\ \nonumber
\end{eqnarray}

Replacing $d$ with $d+2$ in the above equation, we obtain that $p(d+2) = \frac{1}{W} \sum_{t=0}^{W-1} \Pr[|Y_{d+2}| \leq t]$. 
To prove $p(d) > p(d+2)$, we use
the following {\it stochastic ordering} (see Definition\,\ref{def:ordering} below) result established in~\cite{hickey1983majorisation}: 
$|Y_{z}| \le_{st} |Y_{z+2}|$ for any nonnegative integer $z$. 
By Definition\,\ref{def:ordering} , we have $\Pr[|Y_{d}| \leq t] \geq \Pr[|Y_{d+2}| \leq t]$ for $t = 1,2,\cdots,W$. 
When $t = 0$, we have $\Pr[|Y_{d}| \leq t] > \Pr[|Y_{d+2}| \leq t]$
since $\Pr[|Y_{d}|=0] - \Pr[|Y_{d+2}|=0] = \frac{1}{2^{d+1}(d+2)} \binom{d+2}{(d+2)/2} > 0$.
Hence $p(d) = \frac{1}{W} \sum_{t=0}^{W-1} \Pr[|Y_{d}| \leq t] >  \frac{1}{W}  \sum_{t=0}^{W-1} \Pr[|Y_{d+2}| \leq t] = p(d+2)$.
\end{proof}

\begin{definition}\label{def:ordering}
Random variable $X$ is said to be stochastically less than or
equal to random variable $Y$, denoted as $X \le_{st} Y$, if and only if $\Pr[X \leq t] \geq \Pr[Y \leq t]$ for $- \infty < t < \infty$.
\end{definition}

\end{document}